\documentclass[a4paper]{PoS}

\newcommand{\GPDtE}{\langle \tilde{E} \rangle}
\newcommand{\GPDtH}{\langle \tilde{H} \rangle}

\newcommand{\GPDHT}{\langle H_T \rangle}
\newcommand{\GPDETbar}{\langle \bar{E}_T \rangle}

\newcommand{\HT}{\langle H_T \rangle}
\newcommand{\ETbar}{\langle \bar{E}_T \rangle}

\title{Status of the Experimental Studies on DVMP and Transversity GPDs}

\ShortTitle{DVMP and Transversity GPDs}

\author{\speaker{Valery Kubarovsky}\\
        {\normalfont and the CLAS collaboration}\\
        Thomas Jefferson National Accelerator Facility\\
        Newport News, VA 23606,
        USA\\
        E-mail: \email{vpk@jlab.org}}

\abstract{
The cross section of the exclusive   $\pi^0$ and $\eta$ electroproduction reaction $ep\to e^\prime p^\prime \pi^0/\eta$ was measured at Jefferson Lab  with a 5.75-GeV electron beam and the CLAS detector. Differential cross sections $d^4\sigma/dtdQ^2dx_Bd\phi$ and  structure functions $\sigma_U = \sigma_T+\epsilon\sigma_L, \sigma_{TT}$ and $\sigma_{LT}$, as functions of $t$ were obtained over a wide range of $Q^2$ and $x_B$.  
At low $t$, both  $\pi^0$ and $\eta$ are described reasonably well by Generalized Parton Distributions (GPDs)  in which chiral-odd transversity GPDs are dominant. 
Generalized form factors of the transversity GPDs $\HT^{\pi,\eta}$ and $\ETbar^{\pi,\eta}$ were directly extracted from the experimental observables.
The combined $\pi^0$ and $\eta$ data opens the way for the flavor decomposition of the transversity GPDs. 
The first ever demonstration of this decomposition was done for  the transversity GPDs $H_T$ and $\bar E_T$. 
 GPD $\bar E_T$ is connected with 
the density of the polarized quarks in an unpolarized nucleon in the impact parameter space.
 The spin density of  polarized  $u$ and $d$-quarks was evaluated for different values   of Feynman $x$
 from the  GPD model tuned to described the experimental data.
}
\FullConference{23rd International Spin Physics Symposium - SPIN2018 -\\
		10-14 September, 2018\\
		Ferrara, Italy}

\begin{document}

\section{Introduction}	

Understanding nucleon structure in terms of the fundamental degrees of freedom of Quantum Chromodynamics (QCD) is one of the main goals in the theory of strong interactions.   In recent years it became clear that  exclusive reactions may provide information 
about hadron structure
encoded in so-called Generalized Parton Distributions  \cite{Ji,Radyushkin} (GPDs).
For each quark flavor $q$
there are eight GPDs. Four correspond to  parton helicity-conserving (chiral-even) processes,  denoted 
by $H^q$,  $\tilde H^q$,  $E^q$ and  $\tilde E^q$, and 
four correspond to parton helicity-flip (chiral-odd) processes  \cite{ji,diehl},  $H^q_T$,  $\tilde H^q_T$,  $E^q_T$ and  $\tilde E^q_T$. 
The GPDs depend on three kinematic variables: $x$, $\xi$ and $t$. In  a symmetric frame,  $x$ is the   average longitudinal momentum fraction of the struck parton before and after the hard interaction and $\xi$ (skewness) is half of the longitudinal momentum fraction transferred to  the struck parton. The skewness can be expressed in terms of the   Bjorken variable $x_B$  as
$\xi\simeq x_B/(2-x_B)$. Here $x_B=Q^2/(2p\cdot q)$ and $t=(p-p^\prime)^2$, where $p$ and $p^\prime$ are the initial and final four-momenta of the nucleon.

When the theoretical calculations for longitudinal virtual photons were compared with the JLab  $\pi^0$ and $\eta$ data\cite{clas1,clas2,clas3}  
they were  found  to  underestimate the measured cross sections by more than an order of magnitude in their accessible kinematic regions.
The failure to describe the experimental results with quark helicity-conserving operators  stimulated a consideration of the role of the  chiral-odd quark helicity-flip processes. Deeply virtual meson  electroproduction (DVMP), and in particular $\pi^0$ production in the reaction $ep\to e^\prime p^\prime \pi^0$, was 
identified~\cite{Ahmad:2008hp,G-K-09,Goloskokov:2011rd} 
as especially sensitive to the quark helicity-flip subprocesses. 
During the past few years, two parallel theoretical approaches - \cite{Ahmad:2008hp,GL}~(GL) and \cite{G-K-09,Goloskokov:2011rd}~(GK) have been developed utilizing the  chiral-odd GPDs in the calculation of  pseudoscalar meson electroproduction. The GL and GK approaches, though employing different models of  GPDs, lead to {\it transverse} photon amplitudes that are much larger than the longitudinal amplitudes.

\section{Definition of structure functions}

The unpolarized reduced meson cross section is described 
 by 4 structure functions $\sigma_T$, $\sigma_L$, $\sigma_{TT}$ and $\sigma_{LT}$:
\begin{equation}
2\pi\frac{d^2\sigma(\gamma^*p\to p\pi^0)}{dtd\phi_\pi} = 
\frac{d\sigma_T}{dt} + 
\epsilon  \frac{d\sigma_L} {dt}+ 
\epsilon  \frac{d\sigma_{TT}} {dt}  \cos 2\phi+
\sqrt{2\epsilon (1+\epsilon)}   \frac{d\sigma_{LT}} {dt}  \cos \phi.
\end{equation}

\noindent 
References~\cite{Goloskokov:2011rd,GL} obtain the following relations for unpolarized structure functions:

\begin{equation}
\label{SL}
\frac{d\sigma_{L} }{dt}= \frac{4\pi\alpha}{k^\prime}\frac{1}{Q^4}\left\{ \left( 1-\xi^2 \right) \left|\GPDtH\right|^2 -2\xi^2 {Re}\left[ \GPDtH^* \GPDtE \right] - \frac{t^\prime}{4m^2} \xi^2 \left| \GPDtE \right|^2 \right\}
\end{equation}

\begin{equation}
\label{ST}
\frac{d\sigma_{T}}{dt} = \frac{4\pi\alpha}{2k^\prime Q^4} \left[ \left(1-\xi^2\right) \left|\GPDHT\right|^2 - \frac{t'}{8m^2} \left|\GPDETbar\right|^2\right]
\end{equation}

\begin{equation}
\label{SLT}
\frac{d\sigma_{LT}}{dt} = \frac{4\pi\alpha}{\sqrt{2}k^\prime Q^4} \xi \sqrt{1-\xi^2} \frac{\sqrt{-t'}}{2m} { Re} \left[ \langle H_T\rangle^* \langle\tilde{E}\rangle \right]
\end{equation}

\begin{equation}
\label{STT}
\frac{d\sigma_{TT}}{dt} = \frac{4\pi\alpha}{k^\prime Q^4}\frac{t'}{16m^2}\left|\GPDETbar\right|^2
\end{equation}

\noindent 
Here $m$ is the mass of the proton, $t^\prime =t-t_{min}$, where $|t_{min}|$ is the minimum value of $|t|$ corresponding to $\theta_\pi =0$, $k^\prime(Q^2,x_B)$ is a phase space factor and  $\bar E_T = 2\widetilde H_T + E_T$.
The brackets $\langle  H_T \rangle$ and $\langle \bar E_T \rangle$ denote
the convolution of the elementary process 
$\gamma^*q\to q\pi^0$
with the GPDs $H_T$ and $\bar E_T$. They are called them generalized form factors.

\begin{figure*}[t!]
\centering
\includegraphics {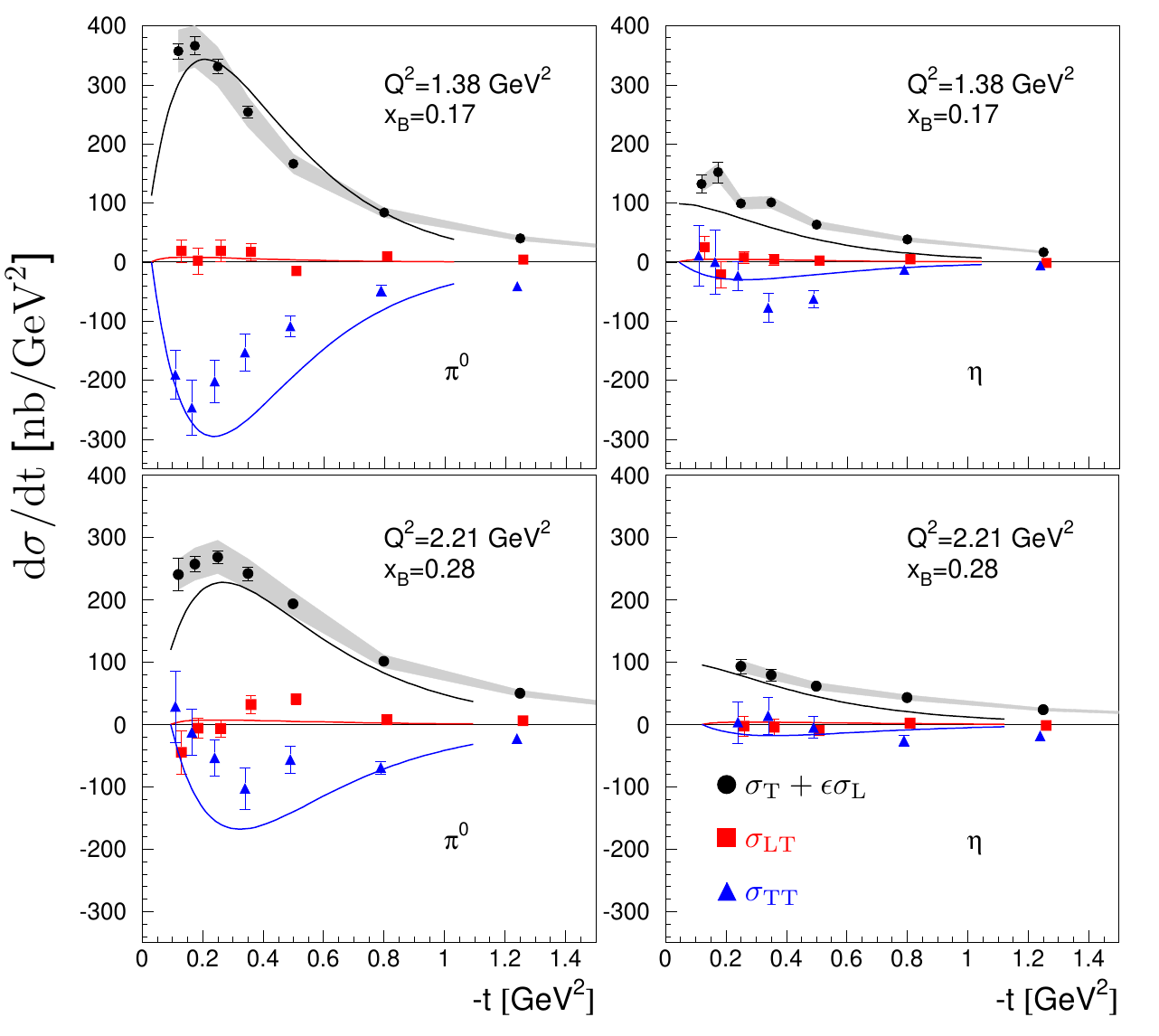}
\caption
{
(Color online) 
The extracted structure functions vs. $t$ for 
the $\pi^0$ (left column) 
and $\eta$ (right column). The top row presents data for the kinematic point  ($Q^2=$1.38~GeV$^2$,$x_B$=0.17)
and bottom row for the kinematic point ($Q^2=$2.21~GeV$^2$,$x_B$=0.28). 
The data and curves are as follows:  
black circles - $d\sigma_U/dt =d\sigma_T/dt +\epsilon d\sigma_L/dt$,
blue triangles - $d\sigma_{TT}/dt$,
red  squares - $d\sigma_{LT}/dt$.
The error bars are statistical only. The gray bands are our estimates of the absolute normalization systematic uncertainties on $d\sigma_U/dt $.
The curves are theoretical predictions produced  with the GPG model of Goloskokov and Kroll~\cite{Goloskokov:2011rd}.
}
\label{fig:structure}
\end{figure*}

\section{Experimental data}
The cross section of the reaction $ep\rightarrow ep(\pi^0/\eta)$ measured by 
the CLAS collaboration at Jlab in bins of $Q^2$, $x_B$, $t$ and $\phi$ were published in 
 Refs.
 \cite{clas1,clas2,clas3}.
Structure functions $\sigma_U=\sigma_T+\epsilon\sigma_L$, $\sigma_{LT}$ and $\sigma_{TT}$ have been extracted from the angular distributions. These functions were compared 
with the predictions of the GPD models \cite{Goloskokov:2011rd,GL}. The result confirmed that the measured unseparated cross sections are much larger than expected from 
leading-twist handbag calculations which are dominated by longitudinal photons. 
As an example, the comparison of the $\pi^0$ and $\eta$ structure functions is shown in Fig.~\ref{fig:structure} for two kinematical bins in $x_B$ and $Q^2$. 
The structure functions
$\sigma_U$ and $\sigma_{TT}$ for $\eta$ are, respectively,  factors of 2.5 and 10 smaller than for $\pi^0$.
However, the GK GPD model \cite{Goloskokov:2011rd}  (curves)  follows the experimental data.
Taken together, the $\pi^0$ and $\eta$ results stringthen 
the statement about the transversity GPD dominance in the pseudoscalar electroproduction process.

\begin{figure}[t!]
\vspace*{-10 mm}
\centerline{
\includegraphics[height=15cm]{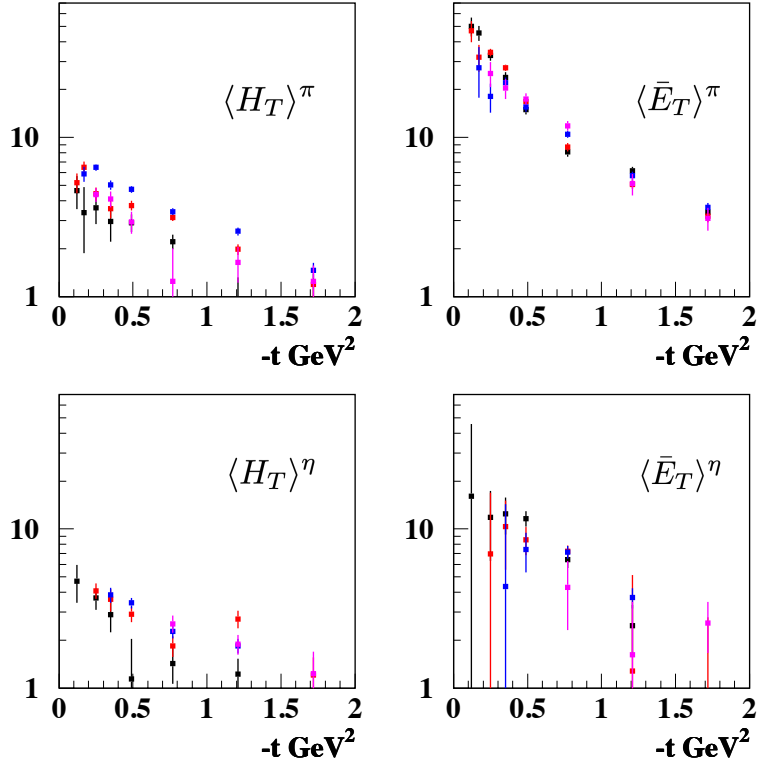}
}
\vspace*{-27mm}
\caption{(Color online) Data points: CLAS.
Top left: $\left|\GPDHT^\pi\right|$,
top right: $\left|\GPDETbar^\pi\right|$,
bottom left: $\left|\GPDHT^\eta\right|$,
bottom right: $\left|\GPDETbar^\eta\right|$ as a function of -$t$ for different kinematics:
($Q^2$=1.2 $GeV^2$,$x_B$=0.15) black, 
($Q^2$=1.8 GeV$^2$,$x_B$=0.22) red, 
($Q^2$=2.2 GeV$^2$,$x_B$=0.29) blue, 
($Q^2$=2.7 GeV$^2$,$x_B$=0.34) magenta. 
}
\label{fig:ht_et_pi0_eta}
\end{figure}


\section{Generalized form factors}
\begin{figure}[t!]
\vspace*{-10 mm}
\centerline{
\includegraphics[width=10cm]{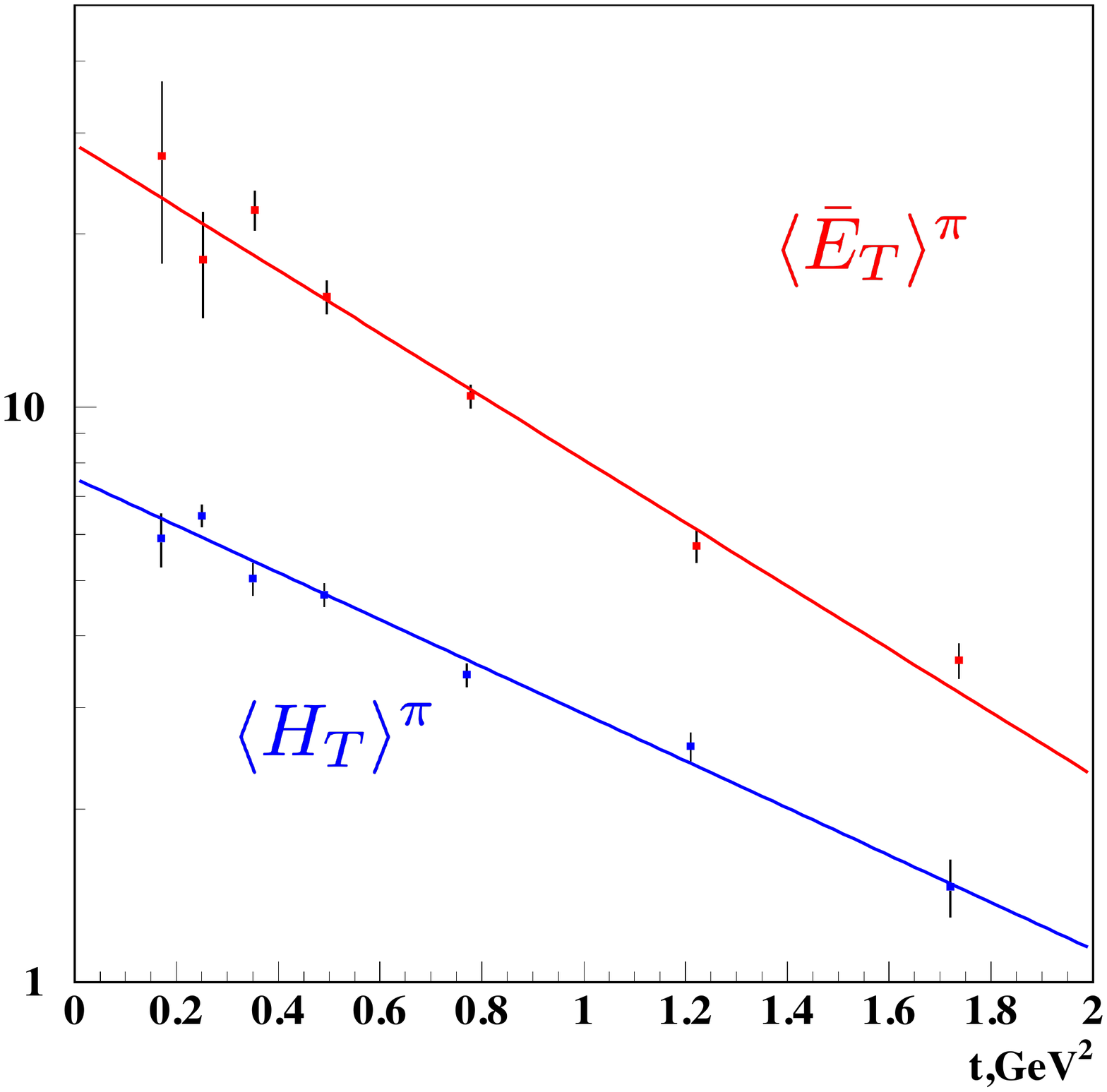}
}
\vspace*{-25 mm}
\caption
{
(Color online) Generalized form factors $|\GPDHT^\pi|$   and $|\GPDETbar^\pi|$
as a function of -$t$ for $Q^2=2.2~GeV^2$  and $x_B$=0.27. 
Top: $|\GPDETbar^\pi|$ in red; 
Bottom: $|\GPDHT^\pi|$ in blue.
}
\label{fig:ht_et_pi0}
\end{figure}

The squared magnitudes of the 
generalized form factors
$\left|\GPDHT\right|^2$ and 
$\left|\GPDETbar\right|^2$  
may be directly extracted from the experimental data (see Eqs. \ref{ST} and \ref{STT} ) in the framework of GPD models.
\begin{equation}\label{ET}
\left|\GPDETbar^{\pi,\eta}\right |^2=\frac{k^\prime Q^4}{4\pi\alpha}     \frac{16m^2}{t'} \frac{d\sigma^{\pi,\eta}_{TT}}{dt} 
\end{equation}
\begin{equation}\label{HT}
\left| \GPDHT^{\pi,\eta}\right|^2=\frac{2k^\prime Q^4}{4\pi\alpha}       \frac{1}{1-\xi^2} \left [ \frac{d\sigma^{\pi,\eta}_{T}}{dt} + \frac{d\sigma^{\pi,\eta}_{TT}}{dt} \right ].
\end{equation}
\noindent
Figure~\ref{fig:ht_et_pi0_eta} presents the modulus of the generalized form factors
$\left|\GPDHT^\pi\right|$, $\left|\GPDETbar^\pi\right|$, $\left|\GPDHT^\eta\right|$ and $\left|\GPDETbar^\eta\right|$
 for 4 different kinematics. 
  Note the dominance of the $\left|\GPDETbar\right|$ over $\left|\GPDHT\right|$ for both $\pi^0$ and $\eta$.
Generalized form factors $\GPDHT^\pi$ and $\GPDETbar^\pi$ are shown in more detail in  Fig.~\ref{fig:ht_et_pi0}. 
The  $\GPDETbar^\pi$ formfactor has steeper t-dependemce than $\GPDHT^\pi$. 
The t-slope parameters, obtained by an exponential fit of the form $e^{bt}$, are 
 $b(\GPDETbar)$=1.27~$GeV^{-2}$ and $b(\GPDHT)$=0.98~$GeV^{-2}$.


\section{Flavor decomposition}

In  electroproduction of $\pi^0$ and $\eta$ mesons the GPDs $F_i$ appears in the following combinations
\begin{equation}
F^\pi_i=\frac{1}{\sqrt{2}}[e_uF^u_i-e_dF^d_i]
\end{equation}
\begin{equation}
F^\eta_i=\frac{1}{\sqrt{6}}[e_uF^u_i+e_dF^d_i-2e_sF^s_i]\\
\end{equation}
\noindent
The $q$ and $\bar q$ GPDs contribute in the quark combinations $F^q_i-F^{\bar q}_i$. 
Hence there is no contribution from the strange quarks if we assume that
$F^s_i\simeq F^{\bar s}_i$. 
For flavor decomposition we have to take into account the decay constants $f_\pi$ and $f_\eta$, the chiral  condensate
constants $\mu_{\pi^0}=$2.57~GeV, $\mu_1=$0.958~GeV and $\mu_8$=2.32~GeV, and the contribution from singlet and octet $\eta$ states \cite{Goloskokov:2011rd}.

\begin{equation}
F^\eta_i=F^{\pi}_i\left(\cos\theta_8-\sqrt{2} \frac{\mu_1}{\mu_8}\frac{f_1}{f_8}\sin\theta_1\right)\frac{f_8}{f_{\pi^0}}\frac{\mu_8}{\mu_{\pi^0}}=\frac{ F^8_i}{k_\eta},
\end{equation}
\noindent
where the mixing angles are:
$\theta_8=-21.2^o$ and $\theta_1=-9.2^o$.
The octet and singlet wave functions are very similar and the decay constants are close as well
$f_8=1.26f_\pi$ and 
$f_1=1.17f_\pi$.
The overall factor for the $\eta$ meson is 
$k_\eta=0.863$.
Using $e_u=\frac{2}{3}$ and $e_d=-\frac{1}{3}$ we will end up with
\begin{equation}\label{decom}
F^\pi_i=\frac{1}{3\sqrt{2}}[2F^u_i+F^d_i]
\end{equation}
\begin{equation}
{k_\eta}{F^\eta_i}=\frac{1}{3\sqrt{6}}[2F^u_i-F^d_i].
\end{equation}
\noindent
Experimentally we have  access only to the $\left|\left<F_i^\pi\right>\right|^2$ and $\left|\left<F_i^\eta\right>\right|^2$
(see Eq.~\ref{ET}-\ref{HT}).
The final equation for the $\left<H_T\right>$ convolution reads

\begin{equation}
\frac{1}{18}       \left | 2\HT^u + \HT^d  \right  |^2 =  \left |\GPDHT^\pi \right |^2\\
\end{equation}
\begin{equation}
 \frac{1}{54}       \left  | 2\HT^u - \HT^d  \right  |^2 = k_\eta^2\left |\GPDHT^\eta \right |^2.
\end{equation}
\noindent
and simular equations for $\ETbar$.

\begin{figure}[t!]
\vspace*{-10 mm}
\centerline{
\includegraphics[width=15cm]{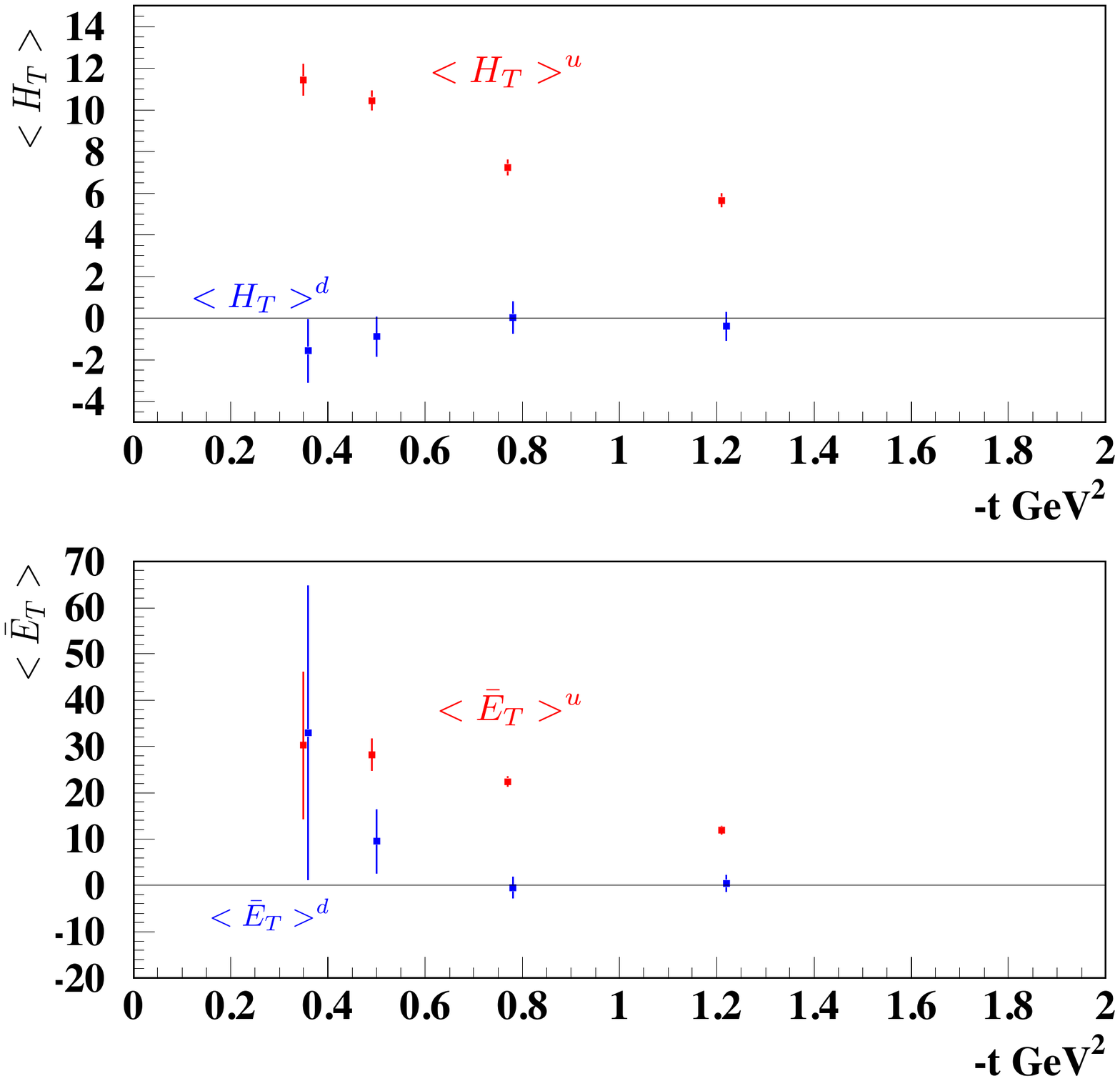}
}
\vspace*{-35 mm}
\caption
{
(Color online) Flavor separated generalized form factors $\GPDHT$ and $\GPDETbar$
as a function of -$t$ for $Q^2=2.2~GeV^2$  and $x_B$=0.27.
Top: $\GPDHT^u$ (red) and $\GPDHT^d$ (blue);
Bottom: $\GPDETbar^u$ (red) and $\GPDETbar^d$ (blue).
}
\label{fig:ud_3}
\end{figure}

\begin{figure*}[t!]
\vspace*{-25 mm}
\centerline{
\includegraphics[width=15cm]{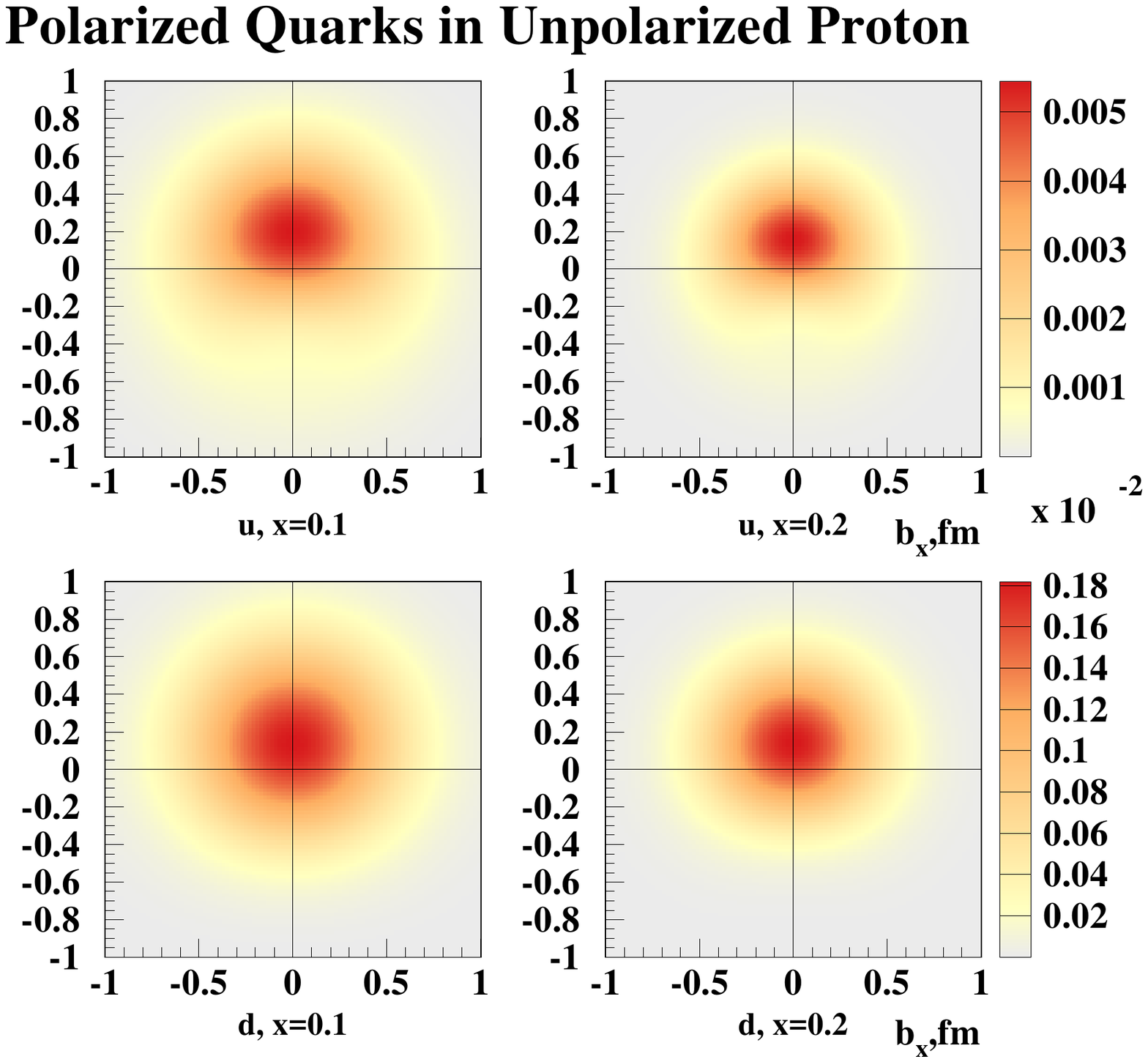}
}
\vspace*{-50 mm}
\caption
{
(Color online) Impact parameter density  of quarks which are transversely 
polarized along $b_x$-axis 
in an unpolarized 
proton.
Top left and right panels are for u-quarks with $x$=0.1 and $x$=0.2, respectively. 
Bottton left and right panels are for d-quarks with $x$=0.1 and $x$=0.2. 
}
\label{fig:den_01-02}
\end{figure*}

The solution of these equations will lead to the flavor decomposition of the generalized form factors 
$\HT^u$ and  $\HT^d$ as well as  $\ETbar^u$ and $\ETbar^d$.
However, the  convolution integrals have real and imaginary parts. So it is impossible to  solve these equations unambiguously with only two equations in hands.
So, in order to estimate the form factors, we make an ad hoc assumption that the
relative phace  $\Delta\phi$ between $\HT^u$ and $\HT^d$  equals 0 or 180 degrees.
Ignoring an overall phase, the form factors are then real, and we arbitrarily choose the solution with 
  $\HT^u$ and  $\ETbar^u$ positive.
Fig.~\ref{fig:ud_3}  presents 
$\HT^u$, $\HT^d$, $\ETbar^u$ and $\ETbar^d$
for one kinematic point $(Q^2=2.2~GeV^2,x_B=0.27)$ calculated with this assumption. 
Note the different signs of $\HT^u$ and  $\HT^d$  and the same sign
of $\ETbar^u$ and $\ETbar^d$.

\section{Quark spin densities in the transverse plane and generalized transversity distributions}
Two-dimensional Fourier transforms of GPDs $H(x,\xi=0,-{\vec \Delta^2)}$ and $\bar E_T(x,0,-{\vec \Delta^2)}$, where $\vec \Delta^2=-t$,
define the spin density of the polarized quarks in an unpolarized proton \cite{Diehl:2005jf}.
\begin{equation}
H(x,{\vec b})=\int \frac{d^2{\vec \Delta}}{(2\pi)^2} e^{-i{\vec b \vec \Delta}} H(x,0,-{\vec \Delta}^2)
\end{equation}
\begin{equation}
\bar E_T(x,{\vec b})=\int \frac{d^2{\vec \Delta}}{(2\pi)^2} e^{-i{\vec b \vec \Delta}} \bar E_T(x,0,-{\vec \Delta}^2)
\end{equation}
\noindent
The GPDs $\bar E_T(x,\xi=0,t)$ and $H(x,\xi=0,t)$ were parametrized in the form \cite{Diehl:2013}
$$
\bar F^q(x,\xi=0,t)=g^q(x) \cdot \exp[(f^q(x)t], 
$$
where $g^q(x)$   and $f^q(x)$ are the GPD forward limit and profile functions respectively.
The Fourier transform for this parametrization reads 

\begin{equation}
\bar F^q(x,b)= \frac{1}{4\pi} \frac{g^q(x)}{f^q(x)}\exp[\frac{-b^2}{4f^q(x)}].
\end{equation}

\noindent For quarks polarized along $b_x$-axis the impact parameter density reads \cite{Diehl:2005jf}

\begin{equation}
\delta^q(x,{\vec b})= \frac{1}{2} [ H^q(x,{\vec b})-\frac{b_y}{m} \frac{\partial}{\partial  b^2}\bar E^q_T(x,{\vec b}) ].
\end{equation}
\noindent
The GPD $H(x,{\vec b})$ describes the density of unpolarized quarks and $\bar E_T(x,{\vec b})$ is related to the distortion of the polarized quark distribution in the transverse plane.
We can map the $u$ and $d$-quark transverse spin density distributions as a function of Feynman $x$ based on the GPD model  \cite{Goloskokov:2011rd} tuned to describe the CLAS data.
For example, Fig.~\ref{fig:den_01-02} shows  the impact parameter density  of transversely 
polarized quarks along the $b_x$-axis  in an unpolarized  proton for  Feynman $x$=0.1 and $x$=0.2.
Note the distortion along $b_y$-axis,  similar for  u and d-quarks. 
Looking at the transverse quark density distribution,
we can say that this width is diminished as $x\to 1$. 
This behavior is typical  for the GPD models.


\section{Conclusion}
Differential cross sections of exclusive $\pi^0$ and $\eta$ electroproduction have been obtained in the few-GeV region at more than 1800 kinematic points in  bins of $Q^2, x_B$, $t$ and $\phi_\pi$. 
Virtual photon structure functions  
$\sigma_U$, $\sigma_{TT}$ and $d\sigma_{LT}$ have been obtained. It is found that $\sigma_U$ and $\sigma_{TT}$ are comparable in magnitude with each other, while $\sigma_{LT}$ is very much smaller than either. 
Generalized form factors of the transversity GPDs $\HT^{\pi,\eta}$ and $\ETbar^{\pi,\eta}$ were directly extracted from the experimental observables
for the first time. It was found that the GPD $\bar E_T$ dominates in  pseudoscalar meson production.
The combined $\pi^0$ and $\eta$ data opens the way for the flavor decomposition of the transversity GPDs. 
Within some simplifying assumptions, the  decomposition has been  demonstrated. 
 The spin density of  polarized  $u$ and $d$-quarks  in the transverse plane was evaluated for different values   of $x$
 from the  GPD model tuned to described the experimental data.


\section*{Acknowledgments}
The author  thanks G. Goldstein, S. Goloskokov, P. Kroll, S. Liuti and A.~Radyushkin  for many informative discussions and making available the results of their calculations. 
This material is based upon work supported by the U.S. Department of Energy, Office of Science, Office of Nuclear Physics under contract DE-AC05-06OR23177.


\end{document}